\documentclass[preprint,12pt]{aastex} 

\begin{document}
\title{Submillimeter Observations of Low Metallicity Galaxy NGC 4214}
\author{Gaku Kiuchi}
\affil{Department of Astronomy, Kyoto University, Kyoto 606-8502, Japan}
\email{gaku@kusastro.kyoto-u.ac.jp}
\author{Kouji Ohta}
\affil{Department of Astronomy, Kyoto University, Kyoto 606-8502, Japan}
\email{ohta@kusastro.kyoto-u.ac.jp}
\author{Marcin Sawicki}
\affil{Dominion Astrophysical Observatory, Herzberg Institute of
Astrophysics, National Research Council, 5071 West Saanich Road, Victoria, B.C., V9E 2E7, Canada}
\email{marcin.sawicki@nrc.gc.ca}
\author{Michael Allen}
\affil{Department of Physics, Washington State University, Pullman, WA, USA 99164-2814}
\email{mlfa@mail.wsu.edu}
\begin{abstract}
 Results of submillimeter ($450 \mu$m and $850\mu$m) observations of 
 a nearby dwarf irregular galaxy NGC 4214 with SCUBA on JCMT are presented.
 We aimed at examining the far-infrared to submillimeter spectral energy 
 distribution (SED) and properties of dust thermal emission 
 in a low metallicity environment by choosing NGC 4214 of which
 gas metallicity (log O/H + 12) is 8.34.
 We found that the SED is quite similar to those of IRAS bright galaxies 
 sample (IBGS) which are local bright star-forming galaxies with 
 metallicity comparable to the solar abundance. 
  A dust temperature and an emissivity index for NGC 4214 obtained by a
 fitting to the single temperature greybody model are $T_{d}$ = $35\ \pm\ 0.8$ K
 and $\beta$ = $1.4\ \pm\ 0.1$, respectively, which are typical values for IBGS. 
 Compiling the previous studies on similar nearby dwarf irregular 
 galaxies, we found that NGC 1569 shows similar results to those of NGC 4214,
 while NGC 4449 and IC 10 SE show different SEDs and  low emissivity indices.
 There seems to be a variety of SEDs among metal poor dwarf irregular
 galaxies. 
 We examined  dependence on intensity of interstellar radiation field
 as well as two-temperature model, but the origin of the difference
 is not clear. 
 Some mechanism(s) other than metallicity and interstellar
 radiation field must be responsible for controlling dust emission
 property.

\end{abstract}

\section{Introduction}
The nature of a dust thermal emission in a low metallicity environment
is presently not well understood.
Bolatto et al. (2000) made a SCUBA (Submillimeter Common Use Bolometer
Array) observation of a nearby low-metallicity (log O/H + 12 $\sim 8.2$,
Skillman et al. 1989) dwarf irregular galaxy 
IC 10 toward its most active star forming region (IC 10 SE)
and found that it has unusual dust properties.
They combined their SCUBA data on IC 10 SE with IRAS (Infrared
Astronomical Satellite), KAO (Kuiper Airborne Observatory), 
and IRAM (Institute de Radio Astronomie Millimeterique) data to derive
the spectral energy distribution (SED) of the dust emission from IC 10 SE.
They could not reproduce the observed SED by the single temperature greybody
model ($S_{\nu}\propto \lambda^{-\beta}B_{\nu}(T_{d})$) with a normal dust 
temperature (e.g., $T_{d}\sim$40 K) and an emissivity index (e.g., $\beta\sim$1.5).
They concluded that the dust emissivity index in IC 10 SE is very 
low ($\beta\sim$0.5) and the dust temperature is high ($T_{d}\sim 55$ K).
Concerning the low gas metallicity of IC 10 SE,
Bolatto et al. (2000) suggest that small grains may be efficiently 
destroyed by intense UV light from OB stars in a low metallicity 
environment, which results in the low emissivity index.

The unusual property of dust emission is also suggested for high 
redshift star-forming galaxies called Lyman break galaxies (LBGs)
 at $z\sim3$. Their rest-frame UV spectra resemble very much those of nearby
star-forming dwarf irregular galaxies like NGC 4214 
(see Figure 1 of Steidel et al. 1996; Shapley et al. 2003), and 
their gas metallicities are estimated  to be log O/H + 12 $\sim$7.8$-$8.7 
(Pettini et al. 2001), which are also comparable to those of
nearby star-forming dwarf irregular galaxies.
Rest-frame UV to optical SEDs indicate that the lights from LBGs are
extinguished by dust to some degree (Meurer et al. 1997; Sawicki and Yee
1998; Shapley et al. 2001; Papovich et al. 2001).
The energy absorbed by dust should be thermally re-emitted in the 
far-infrared (FIR), suggesting that LBGs are detectable in the
submillimeter. Both simple scaling of typical UV$-$FIR SED of nearby
starburst galaxies, as well as more sophisticated energy budget
considerations, predict that LBGs at $z\sim3$ would be detectable 
with SCUBA (Ouchi et al. 1999; Sawicki 2001).
However except for a few LBGs, virtually no LBGs have been detected 
in the submillimeter wavelength (Chapman et al. 2000; Webb et
al. 2003). Ouchi et al. (1999) and Sawicki (2001) suggest the following
possibilities to account for the non-detections:(1) A very high dust emissivity
index in LBGs with the ``normal'' dust temperature of around 40 K or (2)
a very high dust temperature ($T_{d} > 50$ K). In any cases, the SCUBA
non-detections suggest that dust properties in the LBGs may be unusual.
Such high dust temperature is also predicted theoretically for 
an early phase ($<10^7$ yr) of an initial starburst (Hirashita et al. 2002).

Motivated by these results obtained at low-{\it z\ }low-Z and high-{\it
 z\ }low-Z,
 we examine whether the low metallicity 
environment results in unusual emission properties of dust.
It is useful to study SEDs of dust emission in nearby dwarf
irregular galaxies, which provide us a good laboratory for examining 
FIR and submillimeter SEDs and dust emission property 
in the low metallicity environment under an intense star formation. 
Here we present results of submillimeter observations of
the local dwarf irregular galaxies NGC 4214.
NGC 4214 is located at a distance of 4.1 Mpc (Leitherer et al. 1996) and
its gas metallicity is obtained to be log O/H + 12 = 8.34
(Skillman et al. 1989).
Detailed studies such as in CO, HI, and H$\alpha$ were made 
(Taylor et al. 1995; Walter et al. 2001),
and IRAS and KAO data are also available for the object, which
makes it  a good target for our purpose.
In section 2 we describe the observations and data analysis,
and present the resulting maps in submillimeter.
We also present the SED of NGC 4214, and derive a dust temperature
and an emissivity index.
A comparison with nearby massive star-forming galaxies is made
in section 3 and with other local dwarf irregular galaxies in
section 4. Discussion is given in section 5 and summary in section 6.

\section{Observations and results}
NGC 4214 was observed at 450 $\mu$m and 850 $\mu$m simultaneously 
with the SCUBA at JCMT in December 8, 2001 in service mode. 
The observations were made with the jiggle mode resulting in a
map of size $\sim 3^{\prime}$. 
The observed field was centered on the position at 
$\alpha_{2000}=12^h15^m38.9^s$ and $\delta_{2000}=36^{\circ}19^{\prime}49^{\prime\prime}$. 
The total observing time was  1 hour both at 450 $\mu$m and at
850 $\mu$m. A flux calibration was done by 
observing the secondary calibrator CRL618; the assumed flux densities were 
11.2 $\pm$ 1.4 Jy at 450$\mu$m and 4.56 $\pm$ 0.17 Jy at 850 $\mu$m.
The data were reduced by JCMT staff in the standard 
manner using the orac package. A correction for atmospheric opacity was done by
extrapolating the value derived at 225 GHz with Caltech Submillimeter 
Observatory tipping radiometer to 850 $\mu$m and 450 $\mu$m.

\begin{figure}
 \plottwo{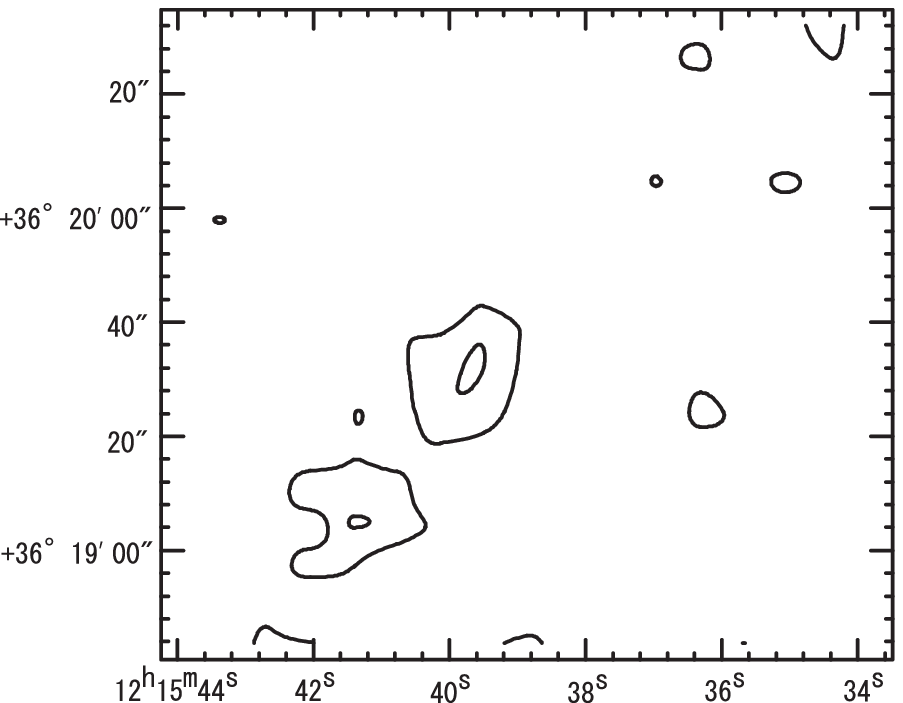}{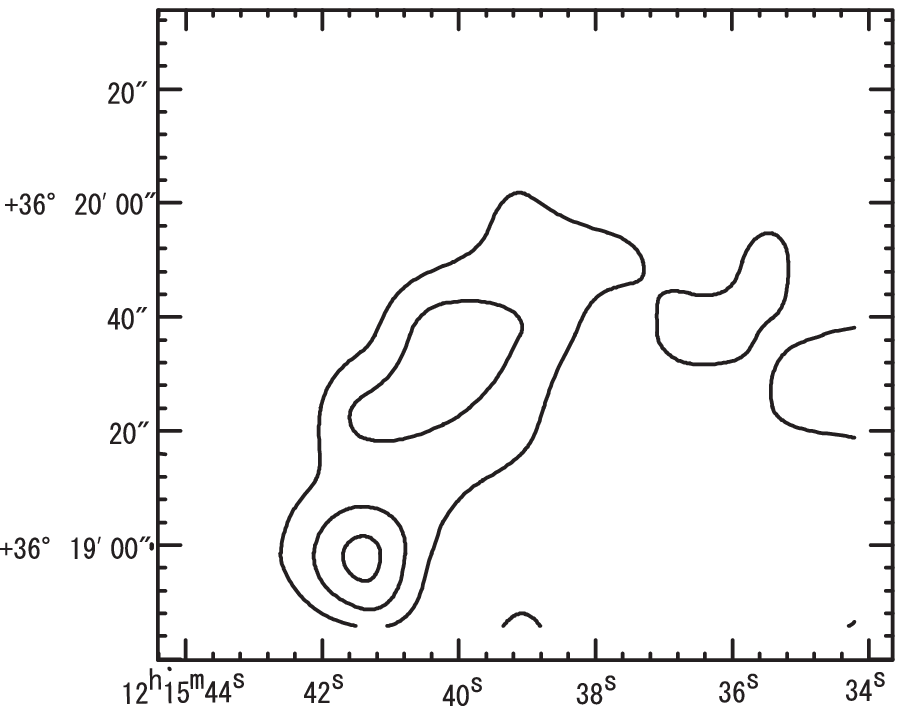}
\caption{Contour maps of NGC 4214 at 450 $\mu$m (left)  and at 850 $\mu$m
 (right).
 Contour levels are at 1$\sigma$, $2\sigma$, and $3\sigma$, where
 $1\sigma=$50 mJy beam$^{-1}$ at $450 \mu$m and $1\sigma=$7 mJy beam$^{-1}$ at $850 \mu$m.
 \label{map}}
\end{figure}

We determined a sky background as a mean of flux values in many
apertures with the same size as the beamsize randomly 
put on a sky region avoiding the object, and subtracted the sky.
Contour maps of NGC 4214 at 450 $\mu$m and 850 $\mu$m after background 
subtraction are shown in Figure \ref{map}. The noises of the two maps
were estimated according to the way of Dunne et
al. (2000) and Dunne \& Eales (2001) by including three kinds of 
noise/error sources; sky subtraction error, 
shot noise,
and calibration error. 
The resulting 1 sigma noises are 50 mJy beam$^{-1}$ and 7 mJy beam$^{-1}$ at 450 $\mu$m
and 850 $\mu$m, respectively.

In Figure \ref{tbl-1}, two peaks are seen in the map at $850\mu$m;
one is located around the center of the map and the other one
$\sim$30$^{\prime\prime}$ southeast of the central component.
Although the significance of the peaks is not so high, 
these peaks almost coincide with those of H$\alpha$ and CO (1$-$0)
distribution (Walter et al. 2001), which makes the detection
reliable.
In the 450 $\mu$m map, the significance of the two peaks is lower, but
the locations are very close to those at 850$\mu$m.
HI and CO (1$-$0) emission are also seen at $\sim$30$^{\prime\prime}$ northwest
of the central component (Walter et al. 2001); the intensity of HI and
CO(1$-$0) is comparable to those at 30$^{\prime\prime}$ southeast. 
However no submillimeter emission and H$\alpha$
emission can be seen. This may indicate that this region is
under a very early phase of star formation and the dust does not
yet form or is not heated up to radiate.
Alternatively, type-II supernova explosions occurred in this region
and destroyed large dust grains (Jones et al. 1996), which 
leads to high temperature not to
be detectable at the longer wavelength.
This subject is interesting with regard of star formation process and
history in NGC 4214, but it is beyond our scope in this paper.
 
To derive the total flux densities at the two bands, we made a
growthcurve in each band adopting the centroid of the flux distribution
as the center.
Resulting total flux densities are
$S_{450} = 1.31 \pm$ 0.56 Jy and $S_{850} = 0.20 \pm$0.04 Jy. 
We examined contamination by thermal free-free emission by extrapolating
the spectrum in the radio wavelength (at 4.85 GHz; Becker et al. 1991)
as $S_{ff}\propto\nu^{-0.1}$, and found that the contribution is 4 \% and 20
\% at 450 $\mu$m and at 850 $\mu$m, respectively.
The CO(3$-$2) emission-line falls into the bandpass at 850 $\mu$m, 
but there is no observation for NGC 4214 at the line. 
Therefore we assume the CO(3$-$2)/CO(1$-$0) ratio to be unity (e.g.,
Lisenfeld et al. 2002 for NGC 1569) to derive the maximum contamination.
With the intensity of CO($1-0$) emission line (Walter et al. 2001),
the estimated maximum contamination of the 
line emission for the flux at 850 $\mu$m is less than 1\%.

\begin{figure}
 \plotone{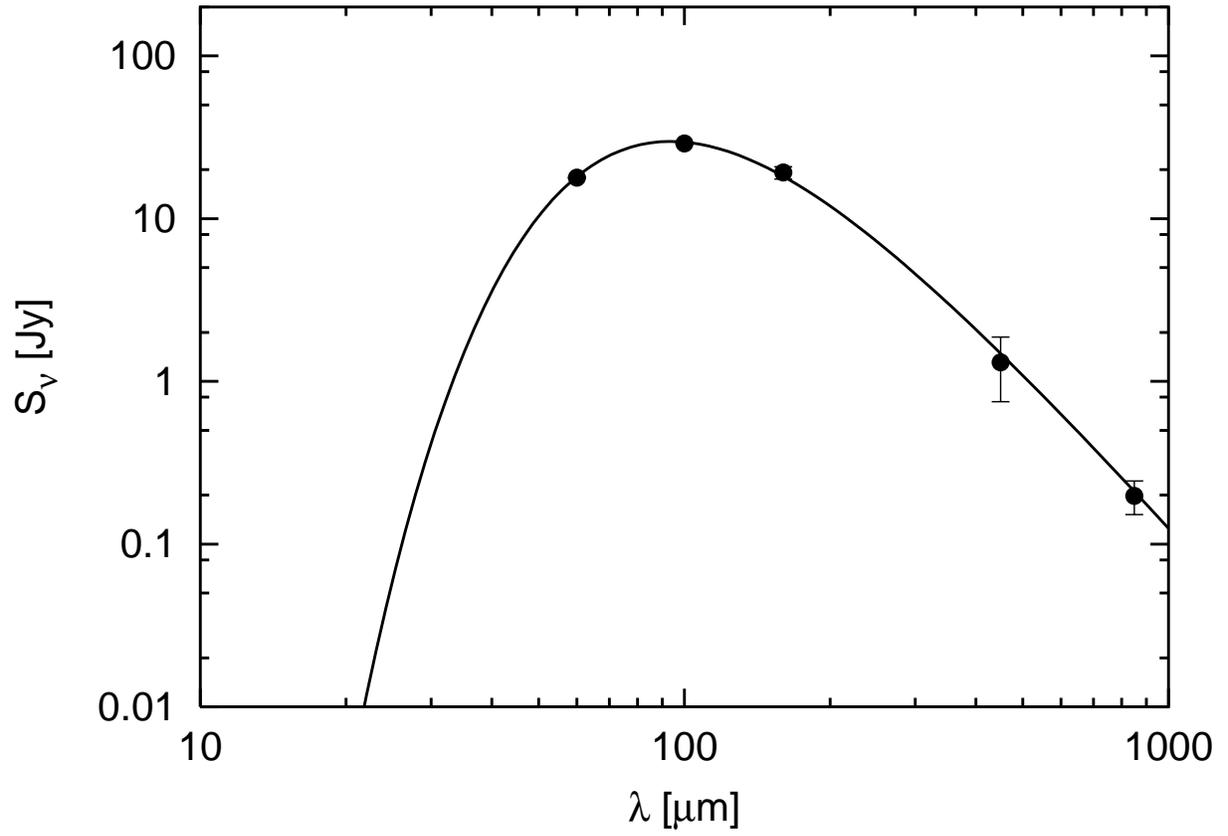}
 \caption{60 $\mu$m to 850 $\mu$m spectral energy distribution of NGC 4214
 (filled circles with an error bar). 
Solid line shows a best fit model with the single temperature
greybody model.\label{SEDs}}
\end{figure}

The SED of NGC 4214 is shown in Figure \ref{SEDs} (the contaminations
are not corrected). Total flux densities
of NGC 4214 at 60 $\mu$m and 100 $\mu$m were obtained with
IRAS by Soifer et al. (1989). The flux at 150 $\mu$m was obtained
with KAO (Thronson et al. (1988)). All the data
are summarized in Table \ref{tbl-1}. 

We fit the resulting SED with a single component temperature model, i.e.,
$S_{\nu} \propto \tau_{\lambda} B_{\nu}(T_{d})$. 
Here $B_{\nu}$ is the Planck
function, $\tau_{\lambda}$ the optical depth with $\tau_{\lambda}
\propto \lambda^{-\beta}$, $T_{d}$ dust temperature and $\beta$ emissivity
index.  The best-fit resulting
parameter values are $T_d =35 \pm 0.8$ K and $\beta = 1.4 \pm 0.1$, and
are given in Table \ref{tbl-2}. The best fit model spectrum is shown in
Figure \ref{SEDs} (solid line). It should be noted that the resulting
parameter values are within a 1 sigma error, even if we correct for the
contaminations.

\section{Comparison with IRAS bright galaxies}
In order to compare the $T_{d}$ and $\beta$ obtained for NGC 4214 with those 
of bright star-forming galaxies,
we derived dust temperatures and emissivity indices of a subset of IRAS
Bright Galaxy Sample (IBGS)(Soifer et al. 1989) by fitting their SEDs
with the single temperature model. 
The flux densities at 60 $\mu$m, and 100 $\mu$m were obtained with 
IRAS, and these at 450 $\mu$m and 850 $\mu$m were obtained with SCUBA by
Dunne \& Eales (2001). 
Although metallicity of these galaxies are not available, since most 
of them are optically bright, they should have high metallicity
(e.g., Skillman et al. 1989).
In fact, for about 20 \% of the IBGS, metallicities are obtained by James et al.
(2002) and most of them lie in 8.6 $\le$ log O/H + 12 $\le$ 9.0. 

\begin{figure}
\plotone{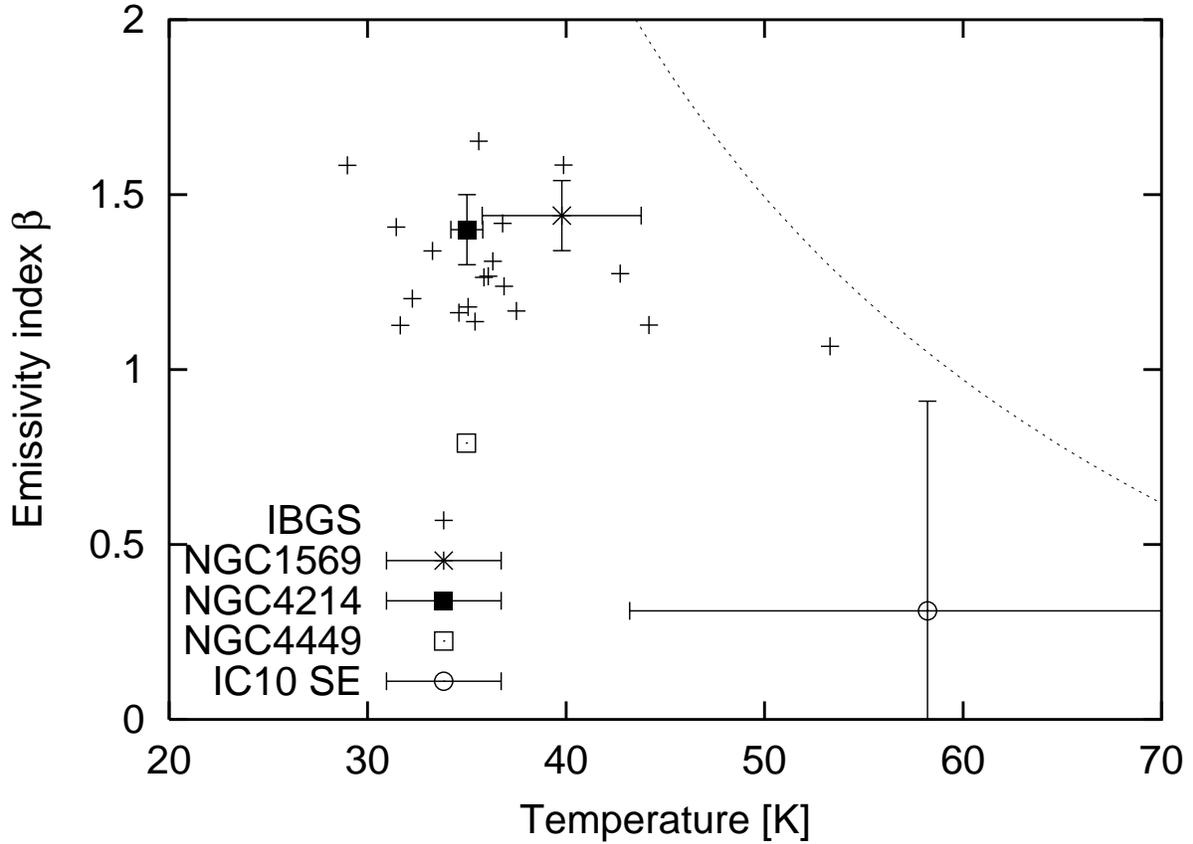}
\caption{Emissivity indices  and dust temperatures derived with the single
 temperature model.  Filled square shows NGC 4214.  Asterisk, open square, and
open circle indicate NGC 1569, NGC 4449, and IC 10 SE, respectively.
Plus signs refer to IRAS Bright Galaxy Sample (IBGS). 
Upper right region above a dotted curve is the maximum
parameter space  allowed for the dust in the LBG cB58 at $z=2.7$
derived from a SCUBA 850 $\mu$m upper limit and UV-optical data
(Sawicki 2001).
\label{T-b}}
\end{figure}

Derived dust temperatures and emissivity indices of the IBGS and NGC
4214 are shown in Figure \ref{T-b}. A typical error for the IBGS sample
is 1 K for a dust temperature and 0.1 for $\beta$.
It should be noted here that some ($\sim$30 \%) of the IBGS are not fitted 
within 1 sigma by using the single dust temperature model. 
Derived dust temperatures of IBGS distribute between 
25 K and 45 K, and emissivity indices distribute between 1.2 and 1.6. 
The dust temperature and emissivity index of NGC 4214 lie
in the region occupied by IBGS. 
Therefore no significant  difference in  FIR-submillimeter
SEDs between NGC 4214 and IBGS depending on metallicity is 
found.

\section{Comparison with NGC 1569, NGC 4449, and IC 10}
In addition to the data of NGC 4214, we compiled FIR$-$submillimeter data from
the previous studies for three  dwarf irregular galaxies having 
similar low metallicity: NGC 1569 (e.g., Lisenfeld et al 2002), NGC 4449
(e.g., B\"{o}ttner et al. 2003) and IC 10 SE (e.g., Bolatto et al. 2001). 
Total flux densities of these at 60 $\mu$m 
and 100 $\mu$m were obtained with IRAS and those at 150 $\mu$m were
 obtained with KAO.
Submillimeter data were obtained with SCUBA on JCMT. 
The data are summarized in Table \ref{tbl-1} with their data sources.
The resulting dust temperatures and emissivity indices obtained
through the SED fitting by using the single temperature model
are listed in Table \ref{tbl-2}.

In Figure \ref{T-b}, the dust temperatures and emissivity indices  of the
three dwarf galaxies are plotted.
As with NGC 4214, NGC 1569 is located in the parameter region 
occupied by IBGS and shows normal nature of the dust emission.
Meanwhile, NGC 4449 and IC 10 SE fall in positions far away from the region
occupied by IBGS, indicating some unusual property of the dust emission.
Their flux densities at 450 $\mu$m and at 850 $\mu$m may be contaminated by
free-free radiations and CO line emissions. 
However even if we correct for these contaminations, the emissivity
index changes within 1 sigma errors, hence 
the contaminations  are not the cause for the large deviations.

\begin{figure}
\plotone{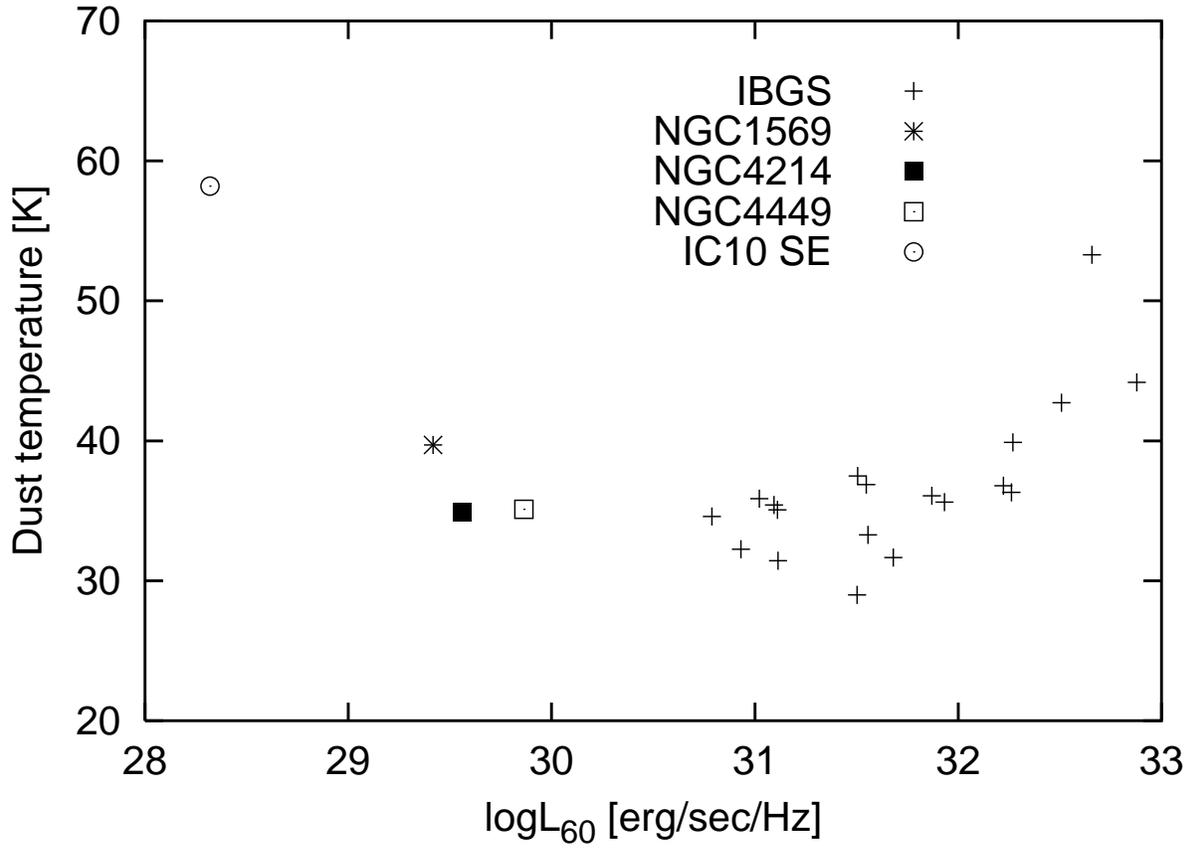}
\caption{Dust temperature against luminosity density at
 60 $\mu$m ($L_{60}$). 
Symbols are the same as those in figure \ref{T-b}. 
$L_{60}$ of IBGS is taken from Dunne \& Eales (2001) using $H_{0}$ = 70
 km s$^{-1}$ Mpc$^{-1}$ and $\Omega_{m}$ = 1 .\label{L60}}

\end{figure}

In Figure \ref{L60}, the dust temperatures are shown against  
the 60 $\mu$m luminosity density $L_{60}$.
The $L_{60}$ of the dwarf irregular galaxies are about two orders of
magnitude smaller than those of IBGS, but the dust temperature
 of them are not lower than those of IBGS. 
If a liner correlation between luminosity-density and temperature exists (Dunne et al. 2000),
it holds so only for bright galaxies and breaks down at 
smaller luminosities.

To summarize,  although the dust temperatures of some dwarf irregular 
galaxies with low metallicity are relatively high
as compared with those of IBGS (majority of which should be metal
rich), the gas metallicity is not the primary parameter controlling
dust temperature and emissivity.
There seems to be a variety in SED among dwarf irregular galaxies with
metallicity of log O/H + 12 $= 8.2 \sim 8.3$.

\section{Discussion}
Although metallicities of the dwarf irregular galaxies studied here
are almost the same, why dust emission properties of IC 10 SE and NGC 4449 
are quite different from NGC 4214 and NGC 1569?. 
Bolatto et al. (2000) suggested that low metallicity and strong 
radiation fields are required to have a low dust emissivity index, which
may lead to high dust temperature. 
If so, IC 10 SE and NGC 4449 should have a stronger
radiation field than NGC 4214 and NGC 1569 which have a normal 
dust emissivity index and dust temperature. 
In order to examine this, we estimate the intensity of interstellar 
radiation field ($ISRF$) for each of them. 
Since IC 10 SE was not observed at UV wavelength, adopting a $U$-band 
luminosity we estimate an average $ISRF$ as $ISRF_{U} =\nu_{U}L_{U}/A$,
where $A$ refers to an area of a dust emitting region in cm$^2$
determined from the 850 $\mu$m contour map. 
(We took the same aperture size as Bolatto et al. (2000) adopted
 for IC 10 SE.)
The $U$-band luminosity $L_{U}$ is derived in the aperture
with the same size as $A$; photometric data are taken from the Lyon-Meudon
Extragalactic Database (LEDA). 
These values are corrected for the Galactic foreground extinction 
(we do not correct for an internal extinction) and are shown in Table
\ref{tbl-3}.
These estimations are expected to have an uncertainty of factor of about
2.
The $ISRF_{U}$s of IC 10  SE and NGC 4449 are comparable to that of NGC 4214,
but are slightly smaller than that of NGC 1569.
A fraction of UV photons radiated from OB stars must be absorbed by dust
 and re-emits from the dust at FIR. 
Thus we also estimate FIR $ISRF$s using 60 $\mu$m fluxes as 
$ISRF_{FIR} = \nu_{60}L_{60}/A$. 
The results are also shown in Table \ref{tbl-3}.
 The $ISRF_{FIR}$ of IC 10 SE is the highest, while that of NGC 4449
is the lowest.
A total $ISRF (= ISRF_{U} + ISRF_{FIR})$ of IC 10 SE is comparable
to that of NGC 1569, rather than that of NGC 4449, for which
the total $ISRF$ is the lowest.
Hence, the high $ISRF$ does not necessarily lead to the 
high dust temperature and low dust emissivity index;
the results indicate the $ISRF$ is not the primary
cause for the difference of SEDs.
Therefore some other mechanism(s) other than metallicity and $ISRF$
must be responsible for controlling dust temperature in these
objects. 

\begin{figure}
\plotone{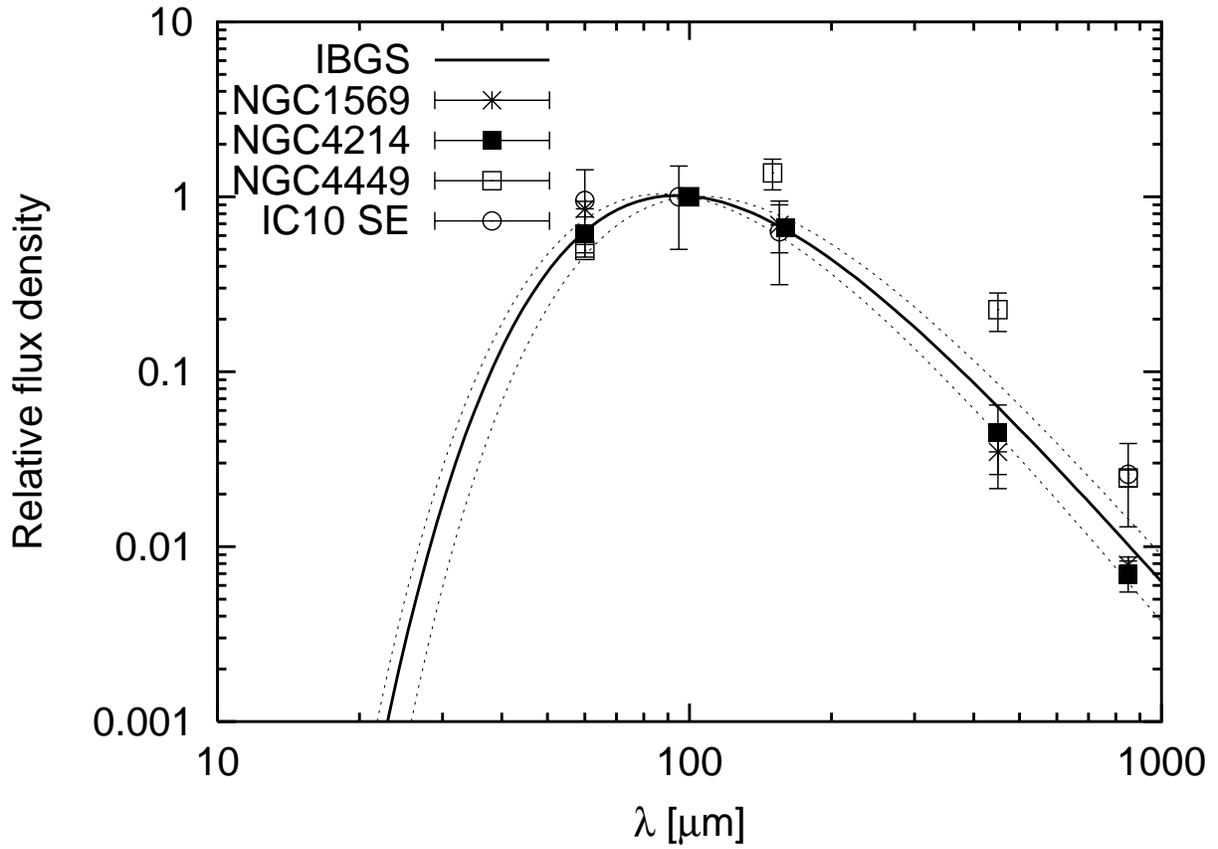}
\caption{Spectral energy distributions normalized at 100$\mu$m.
Symbols are the same as those in figure \ref{T-b}.
Solid line shows the mean SED of IBGS and dot lines represent 
1 sigma deviation from the mean. \label{SEDs_DFG-Dunne}}
\end{figure}

\begin{figure}
\plotone{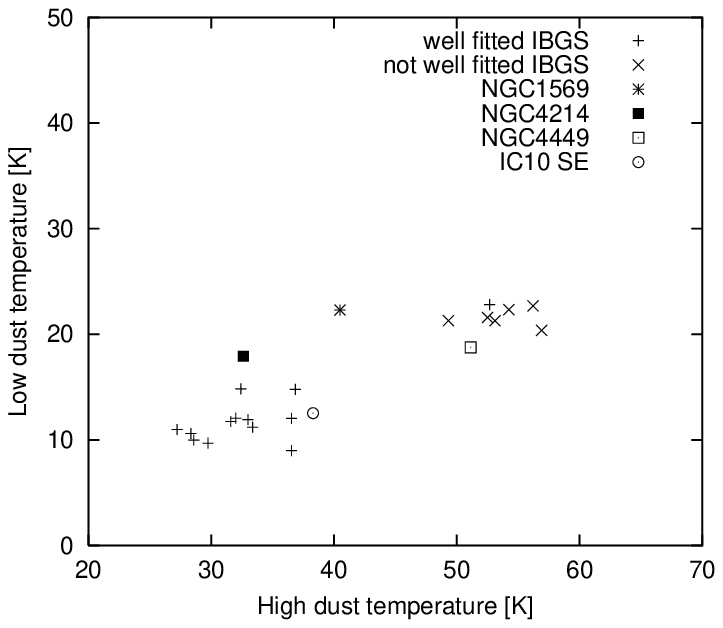}
\caption{High and low dust temperatures derived with the two-temperature
 model. Plus and cross signs refer to galaxies whose SED can be well
 fitted with the single temperature model
 and not well fitted with that, respectively. Other symbols are
 the same as those in figure \ref{T-b}. \label{Twc}}
\end{figure}

Since the parameters used above are derived from the SED fitting 
forced to the simple one-temperature model, we show SEDs of these 
dwarf irregular galaxies normalized at 100 $\mu$m in Figure 
\ref{SEDs_DFG-Dunne} together with a normalized mean SED of IBGS 
to see the shapes of the SEDs in more detail.
The shape of the SEDs of NGC 4214 and NGC 1569 matches the mean SED
of IBGS within its 1 sigma deviation.
The SEDs of NGC 4449 and IC 10 SE, however, have large excesses against 
the mean SED of IBGS in the longer wavelength regime.
The peak of the SED of NGC 4449 locates at the longer wavelength
than 100 $\mu$m.
The SED of NGC 4449 can not be well fitted with the one-temperature
component model  and the uncertainties of derived parameters are
quite large.
Thus we try to make SED fitting by employing 
a two-temperature model with a fixed emissivity index.
When we adopt the emissivity index of 2.0, the distribution of
high-temperature ($T_{\rm h}$) and low-temperature ($T_{\rm l}$) for IBGS shows 
two distinct groups as shown in Figure 6;
one is located around $T_{\rm h} \sim 30$ K and $T_{\rm l} \sim 15$ K
 (group 1),
 while the other around $T_{\rm h} \sim 55$ K and $T_{\rm l} \sim 20$ K
 (group 2).
The galaxies in the group 1  have  cold-to-warm dust mass ratios
of 10--100, while the galaxies in group 2 show the large ($\sim 500-1000$)
dust mass ratios.
The SEDs for the group 1 are well fitted with the one-temperature model, 
while those of group 2 are not well fitted with it.
NGC 4214 and IC 10 SE locate close to group 1, and 
NGC 4449 locates close to group 2. NGC 1569 locates between
the two groups.
This result suggests there is another parameter affecting
the shape of SED other than metallicity and $ISRF$. 
FIR luminosity does not correlate with this trend.
The large mass ratio of cold-to-warm dust may suggest the difference 
of a phase of star formation.
However, when we adopt the emissivity index of 1.5,
the trend seen in Figure \ref{Twc} disappears.
In addition, for some of the IBGS, the obtained high-temperature
is very much close to the low-temperature, indicating that employing 
the two-temperature model is not appropriate for these.
Since the data points presented here is not large for each galaxy and
our data do not have very good signal-to-noise ratios,
getting into detail discussion using a multi-component model is
out of our scope.
More data points and much better signal-to-noise ratios 
seem to be necessary to examine multi-component models
in detail.

\section{Summary}
We observed a nearby dwarf irregular galaxy NGC 4214,
whose gas metallicity (log O/H + 12) is 8.34, 
at $450\mu$m and $850\mu$m with SCUBA on JCMT,
aimed at examining dust emission property in a low metallicity environment. 
The spectral energy distribution in FIR to submillimeter is quite
similar to those of IRAS Bright Galaxy Sample (IBGS), most of which are 
inferred to be metal rich galaxies.
The derived dust temperature and the emissivity index through
the one-temperature model show typical values for IBGS.
Compiling the previous studies on similar nearby dwarf irregular
galaxies, we found that NGC 1569 shows a similar SED and thus 
the dust temperature and the emissivity index to those of NGC 4214,
 while NGC 4449 and IC 10 SE show unusually low emissivity indices, 
and IC 10 SE shows  a significantly high dust temperature. 
There seems to be a variety of SEDs among metal poor dwarf irregular
 galaxies.
We examined  dependence on intensity of interstellar radiation field
 as well as two-temperature model, but the origin of the difference
 is not clear.
Further observations at more wavelengths with a good signal-to-noise
ratios seem to be necessary to discuss the origin of the variety.

We were expecting to find a correlation between dust temperature and
metallicity, but we do not.  This suggests that some other property
other than metallicity and $ISRF$ must be important.
We do not know what that property is at the moment.
One possibility is that the age of the starburst.
Theoretical work (e.g., Hirashita et al. 2002) suggests that
dust temperature is high in the early stage of a galaxy's life and then
drops slowly over time. 
In this scenario, the high dust temperature of cB58 (Sawicki 2001) 
and other LBGs (Ouchi et al. 1999; Chapman et al. 2000) would
suggest that the starbursts in the LBGs are young.  
It would be consistent with the ages ($10-100$ Myr) of stellar 
populations in LBGs obtained through spectral energy distribution
fitting (e.g., Sawicki \& Yee 1998; Papovich et al. 2001).
To better test this hypothesis that dust temperature depends on 
the age of the starburst rather than its metallicity will require 
us to age-date a sample of local galaxies that contains objects 
with a range of metallicities and dust temperatures.

This work is partly supported by a Grant-in-Aid for Scientific Research
(15540233) from the Japan Society for the Promotion of Science.

\clearpage 

\clearpage

\begin{deluxetable}{cccccccc}
\tabletypesize{\scriptsize}
\tablecaption{Flux densities from 60 $\mu$m to 850 $\mu$m. \label{tbl-1}}
\tablewidth{0pt}
\tablehead{
 \colhead{Name} & 
 \colhead{$S_{60}$(Jy)} & 
 \colhead{$S_{100}$(Jy)} &
 \colhead{$S_{150}$(Jy)} & 
 \colhead{$S_{450}$(Jy)} &
 \colhead{$S_{850}$(Jy)} & 
 \colhead{logO/H$+$12\tablenotemark{*}} & 
 \colhead{Ref.}
}
\startdata
 NGC 4214 &$17.9 \pm 0.08$ &$29.0 \pm 0.1$ &$19.2 \pm 1.7$ &$1.31\pm 0.56$
 &$0.20 \pm 0.04$ &8.34 &(1)\\
 NGC 1569 &$44.6\pm 4.5$ &$52.2\pm 5.2$ &$36\pm 11$ &$1.82\pm 0.70$
 &$0.239\pm0.045$  &8.16 &(2)\\
 NGC 4449 &$36\pm 3$ &$73\pm 8$  &$100\pm 20$  &$16.5\pm 4.1$ &$1.8\pm 0.2$  &8.32 &(3)\\
 IC 10 SE &$25.7\pm 12.8$ &$27\pm 13.5$ &$17\pm 8.5$ &$-$ &$0.7\pm 0.35$
 &8.20 &(4)
\enddata

\tablerefs{(1) Soifer et al. 1989 for 60 and 100 $\mu$m,
 Thronson et al. 1988 for 150 $\mu$m, this work for 450 and 850
 $\mu$m; 
 (2) IRAS point source catalog for 60 $\mu$m and 100 $\mu$m,
 Hunter et al. 1989 for 150 $\mu$m, Lisenfeld et al. 2002 for 450
 $\mu$m,  James et al. 2002 for 850 $\mu$m; 
 (3) Hunter et al. 1986 for 60 and 100 $\mu$m, Thronson et al. 1987
 for 150 $\mu$m, B\"{o}ttner et al. 2003 for 450 and 850 $\mu$m; 
 (4) Bolatto et al. 2001 for 60 $\mu$m and 850 $\mu$m, Thronson et
 al. 1990 for 100 $\mu$m and 150 $\mu$m.}
 \tablenotetext{*}{Taken from Skillman et al. 1989.}
  
\end{deluxetable}

\begin{deluxetable}{ccc}
\tabletypesize{\scriptsize}
\tablecaption{Dust temperature and emissivity index. \label{tbl-2}}
\tablewidth{0pt}
\tablehead{
 \colhead{Name} & 
 \colhead{$T_{d}$ (K)} & 
 \colhead{$\beta$}
}

\startdata
 NGC 4214                  & 35.0 $\pm$ 0.8 & 1.4 $\pm$ 0.1 \\
 NGC 1569                  & 39   $\pm$ 4   & 1.4 $\pm$ 0.1 \\
 NGC 4449\tablenotemark{*} & 35             & 0.8 \\ 
 IC 10 SE                  & 58   $\pm$ 15  & 0.3 $\pm$ 0.6
\enddata
\tablenotetext{*}{NGC 4449 is not fitted within 1 sigma with the single
 temperature model.}
\end{deluxetable}

\begin{deluxetable}{ccccccc}
\tabletypesize{\scriptsize}
\tablecaption{Intensity of interstellar radiation fields \label{tbl-3}}
\tablewidth{0pt}
\tablehead{
 \multicolumn{1}{c}{Name} & 
 \multicolumn{1}{c}{distance} & 
 \multicolumn{1}{c}{$\Omega_{source}$} &
 \multicolumn{1}{c}{E(B-V)} &
 \multicolumn{1}{c}{$ISRF_{U}$} &
 \multicolumn{1}{c}{$ISRF_{FIR}$} &
 \multicolumn{1}{c}{Ref.} \\
 \multicolumn{1}{c}{(mag)} &
 \multicolumn{1}{c}{(Mpc)} & 
 \multicolumn{1}{c}{(str)$\times$10$^{-7}$} &
 \multicolumn{1}{c}{} &
 \multicolumn{1}{c}{(erg sec$^{-1}$cm$^{-2}$)} &
 \multicolumn{1}{c}{(erg sec$^{-1}$cm$^{-2}$)} &
 \multicolumn{1}{c}{} 
}

\startdata
 NGC 4214 & 4.1 & 1.7  & 0.022 & 
 1.9$\times$10$^{-2}$ & 6.6 $\times$10$^{-2}$ &(1)\\
 NGC 1569 & 2.2 & 1.92 & 0.56 & 
 5.6$\times$10$^{-2}$ & 1.5$\times$10$^{-1}$ & (2)\\
 NGC 4449 & 4.1 & 5.2 & 0.01 &
 1.3$\times$10$^{-2}$ &4.4$\times$10$^{-2}$ &(3)\\
 IC 10 SE & 0.82 &0.46 & 0.77 & 
 2.5$\times$10$^{-2}$ & 3.5$\times$10$^{-1}$ &(4)
\enddata

\tablerefs{(1) Leitherer et al. 1996 for the distance, Schlegel et al. 1998 for
 E(B-V), de Vaucouleurs 1959 for $U$-band flux;
 (2) Israel 1988 for the distance and E(B-V),
 Kormendy 1977 for $U$-band flux;
 (3) Karachentsev et al. 2003 for the distance, Schlegel et al. 1998 for
 E(B-V), Frueh et al. 1996  for $U$-band flux; 
 (4) Wilson et al. 1996 for the distance, Richer et al. 2001 for E(B-V),
 de Vaucouleurs \& Ables 1965 catalog for $U$-band flux.}
   
\end{deluxetable}

\end{document}